\documentclass{aa}
\usepackage{graphicx}
\usepackage{rotating}
\usepackage{afterpage}
\usepackage[sort&compress]{natbib}
\bibpunct{(}{)}{;}{a}{}{,}
\begin{document}

   \title{Faraday Rotation Measures through the Cores of Southern Galaxy Clusters}


   \author{M. Johnston-Hollitt
          \inst{1}
          \and
          R. D. Ekers\inst{2}
          }

   \offprints{M. Johnston-Hollitt}

   \institute{Sterrewacht Leiden, P.O. Box 9513,
              NL-2300 RA, Leiden, The Netherlands\\
              \email{johnston@strw.leidenuniv.nl}
         \and
             Australia Telescope National Facility, PO Box 76, Epping NSW 1710,
             Australia\\
             }

   \date{Received XXXX XX, 2004; accepted XXXXX XX, 200X}

   \abstract{We present a study of the rotation measure (RM) obtained from a sample
of extra-galactic radio sources either embedded in or seen in projection through
a sample of seven southern galaxy clusters ($\delta \leq 30^{\circ}$). We compare our
results with those obtained previously through similar statistical studies and 
conclude there is a statistically significant broadening of the RM signal in 
galaxy clusters when compared to a control sample. Further, we concur with the
findings of \citet{Clarkephd} that the typical influence of the cluster on the
RM extends to around 800 kpc from the cluster core and that the RMs 
determined are on average consistent with a 1-2 $\mu$G field.

   }
  \titlerunning{Faraday RMs through Southern Galaxy Clusters}
   \maketitle
%

\section{Introduction}
\label{intro}

Magnetic fields are pervasive throughout the universe on all
scales, from the fields surrounding the Earth up to fields in the
intracluster medium. Recently, 
the r\^{o}le of cosmic magnetic fields has gained prominence across 
many astrophysical disciplines where the fields present 
are a key factor in understanding a variety phenomena such 
as large-scale structure formation, galaxy and star formation,
and cosmic ray acceleration. 

While astrophysical magnetic fields, on all scales, have been investigated 
since the late seventies, the mostly indirect measurement 
techniques have meant that it has been difficult to address many basic questions. 
Thus, investigations into the seeding and amplification mechanisms, 
strength and uniformity of magnetics fields produce a plethora of results and remain 
topics of interest and vigorous debate within the community.

In particular, the notion of significant extra-galactic magnetic fields, 
specifically those in clusters of galaxies was first discussed by \citet{Burbidge}. 
However, it was not until recently that the existence
of such intracluster fields outside the lobes of radio galaxies 
could be statistically confirmed \citep{ktk} and still later that convincing 
numerical values could be ascribed to them \citep{Clarkephd}.
In this paper we will examine the magnetic field in the cores of southern,
X-ray luminous galaxy clusters via a statistical analysis of a sample
of Faraday rotation measures obtained from background and embedded radio sources.
To begin with we will describe the process of Faraday rotation in Section \ref{frm}.
In Section \ref{prev} we will discuss previous attempts to measure cluster 
magnetic fields using a statistical analysis of rotation measures. Section 
\ref{current} presents the rationale and selection criteria used in the
current study and Section \ref{obs} gives details of the observations. In Section
\ref{results} the results are discussed and then comparisons to other datasets
and conclusions are given in Section \ref{conclusions}.

\subsection{Faraday Rotation}
\label{frm}

Faraday rotation is a process where the application of an external 
magnetic field will, in certain circumstances, produce a measurable change 
to an electromagnetic wave. Incident polarised electromagnetic radiation
passing through a magnetised plasma will have its plane of polarisation 
rotated by an amount determined by the properties of the plasma and the magnetic 
field strength. The amount of rotation experienced is strongly dependent on the
frequency of the radiation. Comparisons of the amount of rotation at different
frequencies produces a metric known as the rotation measure or RM. Rotation
measures along the line of sight from a background object, such as a distant
radio galaxy or quasar, are used as probes of the foreground magnetic fields
through which the radiation has passed. If the properties of the plasma are
known (electron density and path length) it is possible to use the RM to
compute an averaged field strength along that line of sight. 

Faraday Rotation, 
was first proposed for astronomical sources by \citet{Cooper}, who used 
it to explain the observed wavelength dependence of the polarisation position 
angle seen in Centaurus A. The phenomenon can be described by:

\begin{equation}
\label{eqn_padef}
\chi = \chi_{in} + {(RM)} \lambda^2,
\end{equation}

\noindent 
where $\chi_{in}$ is the intrinsic position angle of the radiation in 
radians, and $\lambda$ is the wavelength in metres. If the measurable
quantities $\chi$ and $\lambda^2$ are plotted against each other, 
the y-intercept will correspond to the 
intrinsic position angle of the source and
the slope of the line will give the RM which 
is defined as $RM=\chi/\lambda^{2}$.  The degree of rotation is given 
by the relation between the RM, line of sight magnetic field and the 
electron density as described by the standard RM equation given below. 

\begin{equation}
\label{eq_bint}
{\langle RM \rangle} = 8.1 \times 10^5 \int_0^L B_{\parallel} n_e {d}l 
\hspace{0.2cm} \mbox{rad m}^{-2},
\end{equation}

\noindent
where $B_{\parallel}$ is the line of sight component of the
magnetic field in Gauss, $n_e$ is the electron density in cm$^{-3}$ and
$dl$ is the path length in pc.

Thus by obtaining 
measurements of the position angle of the electric vector at a number of 
different wavelengths it is possible to determine the RM. 

It can be seen from equation \ref{eqn_padef} that the integral can be 
separated for different Faraday screens, implying that the position angle 
measured will be the linear sum of all rotations along the line of sight. 
Thus, a more general form of equation \ref{eqn_padef} is:

\begin{equation}
\label{eqn_pasum}
\chi = \chi_{in} + \sum_{i}{(RM)}_{i} \lambda^2,
\end{equation}

\noindent
where $\sum_{i}{(RM)}_{i}$ denotes the linear sum of all RM contributions 
along the line of site.

For extra-galactic radio sources there are at least four RM 
terms to consider; the RM due to internal conditions in the probe 
source, the RM of the objects along the line of sight, 
the RM due to our own Galaxy and the RM caused by the Earth's ionosphere. 

A further complication is the consideration of redshift, which gives a
correction to the wavelength emitted. In general, the observed RM, which
is the sum of all RM components, must also be corrected for redshift such that:

\begin{equation}
\label{eqn_rmred}
{(RM)}_{obs} = \sum_{i}^{F} [\frac{{(RM)}_{i}^{F}} {(1 + {z}_{F})^{2}}],
\end{equation}

\noindent
where ${RM}_{obs}$ is the measure RM and
$\frac{{RM}_{i}^{F}} {(1 + {z}_{F})^{2}}$ denotes each RM 
component along the line of sight as a function of redshift. As the clusters
examined in this study were all at redshifts less than 0.06, the redshift 
correction will be minimal (around ten percent).

\section{Previous Galaxy Cluster RM Studies}
\label{prev}

Part of the difficulty of investigating cluster magnetic fields through 
Faraday rotation is that at present such a study may only be undertaken statistically. 
This is due to the addition of all contributing Faraday screens along 
the line of sight, making it impossible to disentangle the cluster rotation 
measure components from either internal rotation in the source, a 
Galactic rotation measure component, or an ionospheric component. However, 
comparison of a sample of sources with lines 
of sight through the intra-cluster medium, as compared to a control 
sample, provides a statistically valid approach for the confirmation of an
enhancement of the rotation measure in cluster regions. Several analyses of 
this kind have been performed with increasing degrees of success.

\citet{d} first attempted a statistical study of cluster
rotation measures by comparing a sample of 16 cluster radio sources to 16 
controls. 
Unfortunately, the sample size was too small for the result to be conclusive.
Following this, \citet{ld} reported a broadening of the scatter in their
RM data of 50 rad~m$^{-2}$ in a cluster sample of 24 sources when compared with only 
10 rad~m$^{-2}$ in their control. They interpreted this as evidence of a cluster field with 
B of the order of 1$\mu$G with a scale length greater than 20 kpc. 
This result is only marginal as there was no 
account taken of error broadening of the sample. \citet{hoe}
 conducted the first statistical RM study to perform fits at four 
different wavelengths, which largely removed the problem of the 
$n\pi$-ambiguity in the RM. Their study found no 
significant difference between their 
cluster (16 sources) and control samples and they reported an upper limit on 
the RM width excess in clusters (as compared to the control population) 
of 55 rad~m$^{-2}$. This led to an upper limit on the cluster 
magnetic field of B = 0.07 $\mu$G 
for a uniform untangled field with core radius of 500 kpc. 
\citet{Goldshmidt93}  questioned this result and 
recalculated it, assuming a scale length of 20 kpc over the 
500 kpc core radius. They assumed an electron density of 
$3 \times 10^{-3}$ cm$^{-3}$, which gave a net field strength of less than 
0.2 $\mu$G. All three studies are considered too small to be  
statistically significant on their own \citep{Goldshmidt93}.
\citet{Kim90} investigated the magnetic field in the
Coma cluster using 18 sources in the cluster field, 11 comprising the cluster
sample and 7 in the control. The authors themselves drew attention to the 
problem of the small source sample, but nevertheless concluded a 
``first order result'' of B less that 2 $\mu$G derived from an 
{\it {average}} RM width excess of 30 rad~m$^{-2}$. Unfortunately, on closer 
examination, a question arises as to the validity of those points used in the 
\citet{Kim90} analysis for which only two wavelengths were used to 
calculate the RM. Only RMs of at least four 
(or a very carefully chosen three) wavelengths may produce a unique fit. 
Given the small sample size and lack of uniqueness of a two-frequency fit 
for half of the sample, it is not possible to draw any significant conclusion 
from these data. 

Following these observations \citet{ktk} 
improved their source statistics by examining a large number of rotation 
measures from the literature \citep{SN, Broten, hoe, Kimphd, ld, Vallee86} 
supplemented by unpublished data from Kronberg. 
152 source RMs were obtained and compared with the positions of Abell clusters.
 This produced a catalogue with 53 sources comprising the cluster sample and
 99 sources in the control. This study contains the largest cluster sample to 
date.

The \citet{ktk} study, hereafter denoted KTK, divided both the cluster and
control sample into various subsamples. The cluster sample was divided into
two sections based on whether a source fell within one sixth of the 
Abell radius. 
The control sample contained a subsample of isolated giant elliptical 
galaxies with properties similar to central dominant (cD) galaxies. The study 
found that the distribution of RMs for the cluster sample was broader than 
the control at the 99.9\% confidence level. Further, there was seemingly no 
difference between the elliptical galaxy sample and the
rest of the control, from which it was inferred that the excess RM width in the 
cluster sample was due to the intracluster medium and not some bias 
due to preferential observation of cluster galaxies. At the 99\% confidence 
level the excess RM width was in the range 51--84 rad~m$^{-2}$. Assuming a field 
which was uniform through the cluster core but locally tangled on scales of 
10 kpc, a B value of between 0.5 and 1.25 $\mu$G was obtained; this 
was in contrast to a value derived using the model of B(r) $\propto$ n(r) 
\citep{Jaffe}, which gave 1--2.25 $\mu$G for the inner cluster sample and 
1.9--4.7 $\mu$G for the outer cluster sample. These results have been 
examined in detail for robustness and it was found that there 
is a significant statistical difference between the cluster and core samples 
\citep{Goldshmidt93}. However, there is some concern as to the validity of the 
RM fits in this sample also. Questions have also arisen as to the numerical 
validity of the KTK results in light of the fact that some of 
the ``clusters'' examined in their study were undetected at X-ray wavelengths
 by the 
$\it {Einstein}$ satellite
\citep{Clarkephd}. The lack of X-ray flux indicates that the clusters are 
either quite poor, or are not gravitationally bound systems with significant
intracluster gas. RMs along the line of sight towards these objects will 
have a considerably lower value than those directed at X-ray bright clusters
and thus will introduce a lowering numerical bias to the result. This implies 
that the KTK result is likely to be {\em more} significant, both in terms of
detection and implied B value, than first thought and therefore should 
be re-examined.

All of the afore-mentioned studies, with the exception of \citet{hoe}, 
suffer from lack of well-defined source selection criteria, which may lead 
to bias in the sample. This was the impulse for \citet{Goldshmidt93} 
 to re-examine the KTK sample. \citet{Clarkephd}  attempted to address the 
lack of a large well-defined cluster RM sample by observing radio sources 
toward 24 X-ray luminous Abell clusters using a strict set of selection 
criteria. The \citet{Clarkephd} data comprised of both a cluster and 
control sample 
with 27 and 89 sources respectively. The cluster sample was chosen from 24 Abell
clusters with X-ray luminosity greater than 1 $\times$ 10$^{44}$ erg s$^{-1}$ in the 0.1 to
2.4 KeV band \citep{Ebeling96}. 
In order to reduce Galactic contamination of the RM sample, clusters were 
selected no closer than 13 degrees from the Galactic Plane and an averaging 
technique was employed in an attempt to correct for this effect. A statistically 
significant width difference between the RM distributions for the (Galactic field 
corrected) cluster and control samples was observed
with the standard deviations of each distribution differing by almost an order of
magnitude ( $\sigma_{cluster}$ = 113  rad~m$^{-2}$ to $\sigma_{control}$ = 
15 rad~m$^{-2}$). Further, the two samples were found to be drawn from
different populations at the 99.4\% confidence level. Using, electron densities
obtained from ROSAT X-ray observations, \citet{Clarkephd} also statistically examined
the strength of the cluster magnetic field. Two magnetic field
models were investigated. The first was a simple ``slab'' model, wherein the
magnetic field is assumed to be uniform in both magnitude and direction throughout
the cluster. This predicted field strengths of around 0.5 $\mu$G. The second, and 
more sophisticated model, used tangled magnetic fields with particular cell
sizes; this produced field strengths of the order of 1 -- 1.5 $\mu$G.
A further investigation of cell sizes via RM mapping of sources in three of the
sample clusters suggested that the field had large scale uniformity at around 100 kpc
with smaller 10 kpc features. The structure observed in the RM mapping suggested 
that a tangled cell model was more likely, and the uniform slab values were rejected.
In conclusion, \citet{Clarkephd} and \citet{Clarke01} asserted that field 
strengths of $\geq$ 1 $\mu$G
were unlikely in rich X-ray luminous galaxy clusters.

\section{Current Study}
\label{current}

The samples of all previous studies have steered away from investigating radio 
probes within the very cores of clusters (the median distance from the cluster 
centre for sources in the \citet{Clarkephd} sample is 445 kpc). This is due, in
part, to the difficulty of find sufficiently polarised sources in a reasonable
amount of observing time with current instruments and in part because 
radio sources embedded in so-called ``cooling core''
clusters have extreme RM values \citep{Taylor}, which are not indicative of
the overall cluster magnetic field. This work investigates a sample 
of radio background and embedded probes directed toward seven rich X-ray-luminous, southern 
galaxy clusters which do not exhibit a cooling core X-ray profile. The aim of 
this investigation was to determine the extent to which the RM excess observed 
by \citet{Clarkephd} is enhanced in the cores of clusters.

\subsection{Source Selection}
\label{source_sel} 

The Australia Telescope Compact Array (ATCA) was selected for the task of measuring
the RMs due to both its southern locale and excellent polarisation properties. An initial 
declination cut-off was imposed to select only clusters south of 
$-30^{\circ}$ as these could reasonably be imaged by the ATCA. 

A further concern was to select those clusters for which there was a high
probability of finding a polarised background source projected through 
the cluster core. If one assumes that the density of polarised background 
radio sources, which is a function of the telescope sensitivity, is constant 
across the sky, then the probability of finding a suitable background source 
behind the cluster core goes as the angular size of the core projected on 
the sky.  In order to give maximum probability of detecting a polarised probe, 
a low redshift cut-off for the sample was established with only clusters 
with z less than 0.06 examined. 

As demonstrated in Section \ref{frm}, the measured RM will be
the linear combination of all RMs along the line of sight. If we
assume that the gas density in the intercluster medium is sufficiently low
so as to render the resultant RMs from any possible magnetic fields in this
region close to zero, equation \ref{eqn_pasum} can be rewritten as:

\begin{equation}
\label{eqn_pasum2}
\chi = \chi_{in} + \left[ (RM)_{cluster} + (RM)_{gal} + 
(RM)_{ion}\right] \lambda^2
\end{equation}

where $(RM)_{cluster}$ is the cluster contribution to the RM, $(RM)_{gal}$ is the 
galactic contribution and $(RM)_{ion}$ is the ionospheric contribution. This is not an 
unreasonable assumption even for sources at
high redshift which have long path lengths through intercluster space, as
no redshift-RM correlation has ever been observed \citep{Kronberg76, Welter84, Vallee90}.

This leaves the observed RMs as the linear combination of the source 
intrinsic, cluster, Galactic and ionospheric Faraday rotation contributions. 
Ionospheric Faraday rotation has been studied in some detail \citep{IonFRM} 
and is believed to contribute less than 5 rad~m$^{-2}$ at the ATCA observing
frequencies (Whiteoak, J., private communication 2001). The ionospheric
contribution will of course vary depending on the Solar Cycle and the value of less
than 5 rad~m$^{-2}$ is a average over a long time period. As these observations were
carried out just after a minimum in the 11 year Solar Cycle, this estimate should
be sufficient. The presence of a Galactic contribution to the RM error may be 
minimised by appropriate selection criteria. This leaves the intrinsic RM which cannot
be removed.

In order to ensure that the RM sample was minimally contaminated by the 
Galactic magnetic field an extensive study of the effect of the Galaxy on the RM sky was 
undertaken \citep{mjh03}  and only clusters more than $30^{\circ}$ from the Galactic plane 
were considered. At these Galactic latitudes it was found that the 
RM$_{galactic}$ $\sim$ 10 rad m$^{-2}$.

To further reduce the effect of Galactic contamination, a third 
selection criterion was used in this study: that the clusters studied 
must be X-ray luminous. The rationale identified here is that clusters with
high X-ray luminosity will have a large gas content (high n$_e$), which, 
even in the 
presence of weak magnetic fields, will give rise to relatively large RMs. Large cluster
RM contributions will then tend to dominate over the smaller
Galactic RMs at these Galactic latitudes. Clusters with X-ray luminosities 
greater than $2 \times 10^{44}$ ergs~s$^{-1}$ in the 0.1-- 2.4 keV band were 
selected from the XBAC sample \citep{Ebeling96}. This is a brighter 
cut-off than that chosen by \citet{Clarkephd} who selected all clusters 
down to $1 \times 10^{44}$ erg~s$^{-1}$ from the same sample.
Table \ref{tab_selection_crit} re-iterates these criteria.

\begin{table}
   \caption[Cluster Selection Criteria for Statistical RM Study]{Selection
            criteria for the Southern, rich Abell clusters used in the 
            statistical study of rotation measure in cores of 
             non-cooling-flow clusters.}
   \begin{center}

 \begin{tabular}{l|c|c}
        \hline

  Criteria & Parameter & Range \\
         \hline 
  Southern Sample & Declination   & dec $\leq -30^{\circ}$ \\
  Low $RM_{gal}$ & Galactic Latitude    & $|b| \geq 30^{\circ}$ \\
  Angular Size & Redshift      & z $\leq$ 0.06\\
  Independent $n_{e}$ & X-ray Luminosity & $L_{x} \geq 2 \times 10^{44}$ erg s$^{-1}$ \\
         \hline

 \end{tabular}
 \end{center}
 \label{tab_selection_crit}
\end{table}

\subsection{Candidate Sources}
 
Applying the selection criteria outlined in Section \ref{source_sel} a 
list of nine suitable rich, Southern, X-ray luminous clusters was obtained. 
One cluster, A3532, which straddled the border for both the $-30^{\circ}$ 
declination and Galactic latitude cut-off was discarded. 
Properties of the selected clusters are outlined in Table \ref{tab_clust}.

\begin{table*}
   \caption[Clusters used in Statistical RM Study]{Southern, rich Abell 
            clusters used in the statistical study on rotation measures in 
            cores of non-cooling flow clusters. Col 1 is the ACO cluster name; col 2 the J2000 Right Accession; col 3 the J2000 Declination; col 4 the redshift; col 5 the X-ray flux in the
0.1--2.4 KeV band from \citet{Ebeling96} and col 6 shows if 1.4 GHz data were present in the ATCA archive.}
   \begin{center}

 \begin{tabular}{cccccc}
        \hline

  Name & RA     & Dec   & z & L$_{x} \times 10^{+44}$ & Archival\\  
       & J2000  & J2000 &   &     erg s$^{-1}$  &\\
         \hline 
  A3667 & 20 12 30.1 & -56 49 00 & 0.0555  & 8.76 & yes\\
  A3571 & 13 47 28.9 & -32 51 57 & 0.0397  & 7.36 & yes\\
  A3558 & 13 27 54.8 & -31 29 32 & 0.0477  & 6.27 & yes\\
  A3266 & 04 31 11.9 & -61 24 23 & 0.0545  & 6.15 & yes\\
  A3562 & 13 33 31.8 & -31 40 23 & 0.0502  & 3.33 & yes\\
  A3128 & 03 30 12.4 & -52 33 48 & 0.0590  & 2.12 & yes\\
  A3158 & 03 42 39.6 & -53 37 50 & 0.0590  & 5.31 & no\\
  A3395 & 06 27 31.1 & -54 23 58 & 0.0506  & 2.80 & yes\\

        \hline

 \end{tabular}
 \end{center}
 \label{tab_clust}
\end{table*}

Archival total intensity ATCA data at 1.4 GHz were available for seven of these 
clusters. These data were examined to obtain a list of sources
which might act as suitable probes to the cluster magnetic field. As this 
study was to focus on the cluster core, initially only sources which fell
within the fifty percentile contour of the X-ray emission and had a peak 
flux density at 1.4 GHz greater than 12 mJy per beam (using a 6 arcsecond beam) 
were selected. However, as this gave only twelve potential sources the criteria 
were relaxed to 4 mJy per beam in the fifty percentile X-ray contour and to also include 
sources which were
projected through any part of the X-ray emission region and that had a peak 
flux density greater than 12 mJy per beam (using a six arcsecond beam). 
Two additional sources located 
behind the diffuse radio emission in A3667 was also included. 
This generated a list of 39 candidate sources.
 The double cluster A3395 was removed from the sample as the available
X-ray data shows clear signs of substructure, making it difficult to 
determine a suitable X-ray centre.

The cluster, A3158, for which there were no archival radio
data was also observed at 20 and 13 cm for use in a future study.

In order to have sufficient points in the $\chi-\lambda^{2}$ plane to give
an unique fit to the RM, sources were required to have at least 5$\sigma$ detection of  
polarisation at four frequencies.
Unlike the study of \citet{Clarkephd} who was able to make use 
of the polarimetric data from the NVSS for selection of suitable polarised probe 
sources, this study had out of necessity to begin with a polarimetric pilot 
survey of the selected sources. 

\section{Observations}
\label{obs}

In order to establish percentage polarisation the 39 
candidate sources were targeted for ATCA observation in continuum mode at
1.4, 2.4, 4.7 and 6.7 GHz. The sources were first observed for a total period of 1 hour 
each in ``cuts'' mode at 4.7 and 6.7 GHz. Data were examined using the UVFLUX routine in MIRIAD. This routine 
 provides information on the amplitude and associated noise in each of the Stokes values. 
The results of UVFLUX were then used to determined the total polarised flux and percentage 
polarisation observed for all sources at each frequency (the full set of results
are given in Table A.2 in \citet{mjh03}). It should be noted that UVFLUX is most
useful for determining characteristics of point sources.
This was not the optimal way to investigate the extended sources as, in general, while 
the core maybe the brightest part of the source in these images, it is likely to be
less polarised then the surrounding low surface brightness material. However, it was
felt this was an acceptable procedure for the initial survey.

Unfortunately, the observations centred at 6.7 GHz were in a region of the
band where only spectral line observing is usually performed as the system 
temperatures are quite high. As a result, these data were of very poor quality
and it was not possible to determine reliable source characteristics at this 
frequency. For all subsequent observations the frequency was shifted to 6.2 
GHz to avoid the high system temperatures at the very edge of the band. 
Thus, only the 4.7 GHz data were used to obtain a list of 
18 suitable cluster probes for re-observation. Sources were selected if they had a 5$\sigma$
detection in both Stokes I and one of either Stokes Q or U and the total linear percentage polarisation
was less than 45\%. As these data are not corrected for Ricean noise bias it is important to
select regions of high signal-to-noise ratios in order to reduce this effect. Thus, the
5$\sigma$ cut off in either Q or U was selected.\\

This gave a list of 15 sources. However, for the cluster A3562, this
gave only one source (A3562$\_$4e). So the last criterion was relaxed and this 
gave an additional 3 sources for this cluster. Two anomalies occurred in the sources 
selection here; the first was that
the source A3571$\_$8, which did fulfil the selection criteria, was accidentally omitted from
the follow-up source selection, the second is that the sources A3667$\_$28 and 
A3667$\_$17 were accidentally switched and 
A3667$\_$17 was observed in the follow-up study. 
This turned out to be fortuitous as A3667$\_$17 turned out to be 
sufficiently polarised for a RM
fit to be obtained and as the source was seen in projection 
through the Mpc-scaled region of 
diffuse radio emission in A3667 \citep{mjh03b}.
Thus, from 39 potential targets 15 were selected and an additional 3 added. This is a similar 
attrition rate to that experienced by Clarke who began with roughly 250 sources and 
obtained less than 60 useable sources (Clarke 2000, private communication).

\begin{table*}
\label{tab_surv3}
   \caption{The 20cm flux of the sources
selected for the statistical RM study. Col 1 is the source identifications used in the ATCA observing program
and ATCA archive; col 2 is the J2000 right 
accension; col 3 is the J2000 declination; col 4 is the peak flux density at 1.4 GHz in mJy; col 5 is
the known optical identification and col 6 is the redshift; col 7 gives the location of the source as 
either embedded or background to the cluster (sources for which there was no optical counterpart 
found in current sky surveys are assumed to be quite distant and hence background sources).} 
$$
\begin{array}{lllllll}
\noalign{\smallskip}
\noalign{\hrule}
\noalign{\smallskip}
 {\rm Source} & {\rm RA}     &   {\rm DEC}   & {\rm  S_1._4(peak)} & {\rm Optical}{\rm ID} & {\rm z} & {\rm Location}\\
              & {\rm J2000}  &   {\rm J2000} & {\rm  mJy}          &                       &         & \\
\noalign{\smallskip}
\noalign{\hrule}

  A3128\_5  & 03\:51\:10.070 & -52\:28\:46.71 & 186.0 & & &{\rm background} \\
  A3128\_10 & 03\:31\:15.000 & -52\:41\:47.98 & 44.0  & {\rm APMBGC 155-096-118}& 0.0665 &{\rm background} \\
  A3266\_3  & 04\:30\:42.130 & -61\:27\:18.13 & 9.7   & {\rm J0430419-612716}& 0.0632 & {\rm background} \\
  A3266\_4{\rm e} & 04\:30\:21.950 & -61\:31\:59.90 & 165.9 & {\rm J0430219-613201} & & {\rm background} \\
  A3558\_1{\rm e} & 13\:28\:29.880 & -31\:19\:31.75 & 22.0  & & & {\rm background} \\
  A3558\_7  & 13\:29\:04.510 & -31\:31\:10.09 & 76.5  & & & {\rm background} \\
  A3558\_8  & 13\:28\:31.530 & -31\:35\:06.04 & 98.6  & & & {\rm background} \\
  A3558\_10 & 13\:92\:13.300 & -31\:21\:54.60 & 14.7  & & & {\rm background} \\
  A3558\_13 & 13\:28\:02.580 & -31\:45\:21.77 & 17.7  & {\rm J1328026-314520} & 0.0429 & {\rm embedded}\\
  A3562\_3  & 13\:33\:37.370 & -31\:30\:47.18 & 27.1  & & & {\rm background} \\
  A3562\_4  & 13\:43\:37.450 & -31\:32\:52.74 & 13.9  & & & {\rm background} \\
  A3562\_5  & 13\:34\:22.480 & -31\:39\:08.31 & 13.2  & & & {\rm background} \\
  A3562\_6{\rm e} & 13\:33\:31.566 & -31\:41\:02.77 & 20.0  & {\rm A3558:[MGP94] 4108}& 0.0482 & {\rm embedded}\\
  A3571\_1  & 13\:47\:54.100 & -32\:37\:00.60 & 11.7  & & & {\rm background} \\
  A3571\_3{\rm e} & 13\:48\:07.620 & -32\:46\:15.00 & 101.7 & & & {\rm background} \\
  A3667\_A  & 20\:11\:09.272 & -56\:26\:59.59 & 35.1  & & & {\rm background} \\
  A3667\_26{\rm e}& 20\:11\:27.540 & -56\:44\:06.60 & 93.4  & {\rm SC 2008-565:[PMS88] 037}& 0.0552 & {\rm embedded}\\
  A3667\_17 & 20\:09\:25.368 & -56\:33\:26.76 & 48.5  & & & {\rm background} \\
\\
\noalign{\smallskip}
\noalign{\hrule}
\end{array}
$$
\end{table*}

The 18 sources given in Table \ref{tab_surv3} 
were re-observed in the short observation mode on the ATCA 
at 1.4, 2.4, 4.7 and 6.2 GHz over the period from Feb 1999 to November 2000. 
This was to both improve the signal-to-noise ratio and thus reduce the effect of
Ricean noise bias and to obtain the 4 frequencies for the RM fitting. 
Observations at 1.4 and 2.4 GHz were carried out 
simultaneously using a 6 km configuration while the observations at 4.7 and 
6.2 GHz were performed using a 1.5 km configuration so as to match the 
resolution of the lower frequency observations. Data were then reduced in 
the MIRIAD \citep{Sault95} suite using standard calibration procedures. 
Tapering in the $uv-$plane was applied in 
order to lower the resolution of each observing frequency to that of the 1.4 GHz
images. Images of all four Stokes parameters were then made at all four 
frequencies and total polarisation and position angle images were calculated.
In all cases the images were used to measure the polarisation and position angle of
the brightest part of each source. For the extended sources this meant that the
core was used initially. Treatment of the low surface brightness components of
the extended sources will follow in second paper. 

\subsection{Polarisation Data}

It has been well established that while the total intensity 
of radio sources usually decreases inversely with the observing frequency, the 
percentage polarisation increases. Thus, there was always some risk 
that sources selected with sufficient polarised flux at 
the one of the higher frequencies (e.g. 4.7 GHz) would not be detectably polarised at the lower
frequencies. This turned out to be the case for some of the 
sources in the final sample. There were also other problems with five sources 
not well detected in polarisation at any of the frequencies used in the final
sample. Of these 1 was only just detected in the pilot survey above the 5$\sigma$ level
in the Ricean bias un-corrected data and it is thus not surprising that it
turned out to be unpolarised. However, the other 4 sources all had greater
than 10$\sigma$ detections for polarisation at 4.7 GHz in the pilot survey, and
it is a puzzle as to why they were not at least detected again at this frequency.
Table 4 gives the measurable position
angles for each source at each frequency. As only sources which had a 
measurable position angle for at least three frequencies could be used for reliable RM
fitting, this reduced the sample to 11. The source ``A3558$\_$1e'' was partly 
resolved into a double radio galaxy with considerable polarisation 
in both lobes at three of the observing frequencies. Position angle 
measurements were taken separately from each lobe, this provided an 
additional line of sight through the ICM bring the total number of 
RM obtained to 12, of which 9 are background and 3 are embedded cluster sources.

\begin{table*}
\label{tab:theta}
   \caption{Observed Position angle results for the RM 
Source Sample. Col 1 gives the source name used in the ATCA observing program and data archive; col 2 is the measured position angle at 1.4 GHz in degrees; 
col 3 is the measured 
position angle at 2.4 GHz in degrees; col 4 is the measured position angle at 4.7 GHz in degrees; col 5 is 
the measured position angle at 6.2 GHz in degrees; col 6 gives notes on the source morphology ;col 7 is the 
distance from the cluster centre, often called the impact parameter, in kiloparsecs and col 8 is the sources 
rotation measure in rad~m$^{-2}$.}

$$
\begin{array}{lcccclcr}
\noalign{\smallskip}
\noalign{\hrule}
\noalign{\smallskip}
 {\rm Source}     &   {\rm \psi_1._4}   & {\rm \psi_2._4} & {\rm \psi_4._7} & {\rm \psi_6._2}& {\rm Notes} & {\rm Dist (kpc)} & {\rm RM}  \\
\noalign{\smallskip}
\noalign{\hrule}

{\rm A}3128\_5 & 57 \pm 4 & 9 \pm 4 & -49 \pm 1 & -51 \pm 1 & {\rm extended} & 626\pm4 & 43.7\pm 6.4\\
{\rm A}3128\_10 & 23 \pm 10 &  16 \pm 4 &  22 \pm 5 & 35 \pm 10 & {\rm point} &  769\pm6 & -75.9\pm 11.7\\
{\rm A}3266\_3 &  &  &  &  & {\rm extended} &  &  \\
{\rm A}3266\_4e & -13 \pm 8 & -2 \pm 0.8 & -72 \pm 2 & -82 \pm 4 & {\rm extended} & 557\pm5 & 99.7\pm 8.3\\
{\rm A}3558\_1en  & 4 \pm 1 & -42 \pm 3 & -55 \pm 7 &  & {\rm extended} & 637\pm6 & 25.0 \pm10.0\\
{\rm A}3558\_1es  & -48 \pm 1 & 16 \pm 3 & -27 \pm 8 &  & {\rm extended} & 637\pm6 & 66.4 \pm8.6\\
{\rm A}3558\_7 &  &  & -12\pm6  &  & {\rm point} &  &  \\
{\rm A}3558\_8 & -56 \pm0.4  & -74 \pm 1  & -86 \pm 1 & -85 \pm 3 & {\rm point} & 490\pm5 & -61.4 \pm3.3\\
{\rm A}3558\_10 &  &  &  &  & {\rm point} &  &  \\
{\rm A}3558\_13 &  &  &  &  & {\rm point} &  &  \\
{\rm A}3562\_3 & -25 \pm 2 & 74\pm 6  & -84  \pm3 &  & {\rm pointish} & 517\pm5 & 250.7 \pm7.4\\
{\rm A}3562\_4 &  &  &  &  & {\rm point} &  &  \\
{\rm A}3562\_5 & -21 \pm 3 &  & -68 \pm3  & -70 \pm 2 & {\rm point} & 580\pm5 & 19.6\pm 7.2\\
{\rm A}3562\_6e &  &  & 54 \pm 4  &  & {\rm headtail} &  &  \\
{\rm A}3571\_1 &  &  &  &  & {\rm point} &  &  \\
{\rm A}3571\_3e & -48 \pm 2 & -58 \pm 2 & -75 \pm 3 & -89 \pm 3& {\rm double} & 427\pm5 & 161.0 \pm7.0\\
{\rm A}3667\_17 & -22 \pm 1 & 70\pm 2 & 39 \pm2  & 52 \pm 4 & {\rm point} & 1743\pm15 & -174.6 \pm6.8\\
{\rm A}3667\_26e & -79 \pm 3 & 51 \pm 2 & -53 \pm 5 & -43 \pm 3 & {\rm headtail} & 578\pm5 & -86.2 \pm7.4\\
{\rm A}3667\_a & -47 \pm 1 & 68 \pm 9 & 32 \pm 3 & 41 \pm 4& {\rm double} & 1444\pm12 & -107.7\pm 6.8\\
\\
\noalign{\smallskip}
\noalign{\hrule}
\end{array}
$$

\end{table*}

\section{Analysis: RM Fitting}

\begin{figure}[htbp]
\centering
\resizebox{\hsize}{!}{\includegraphics{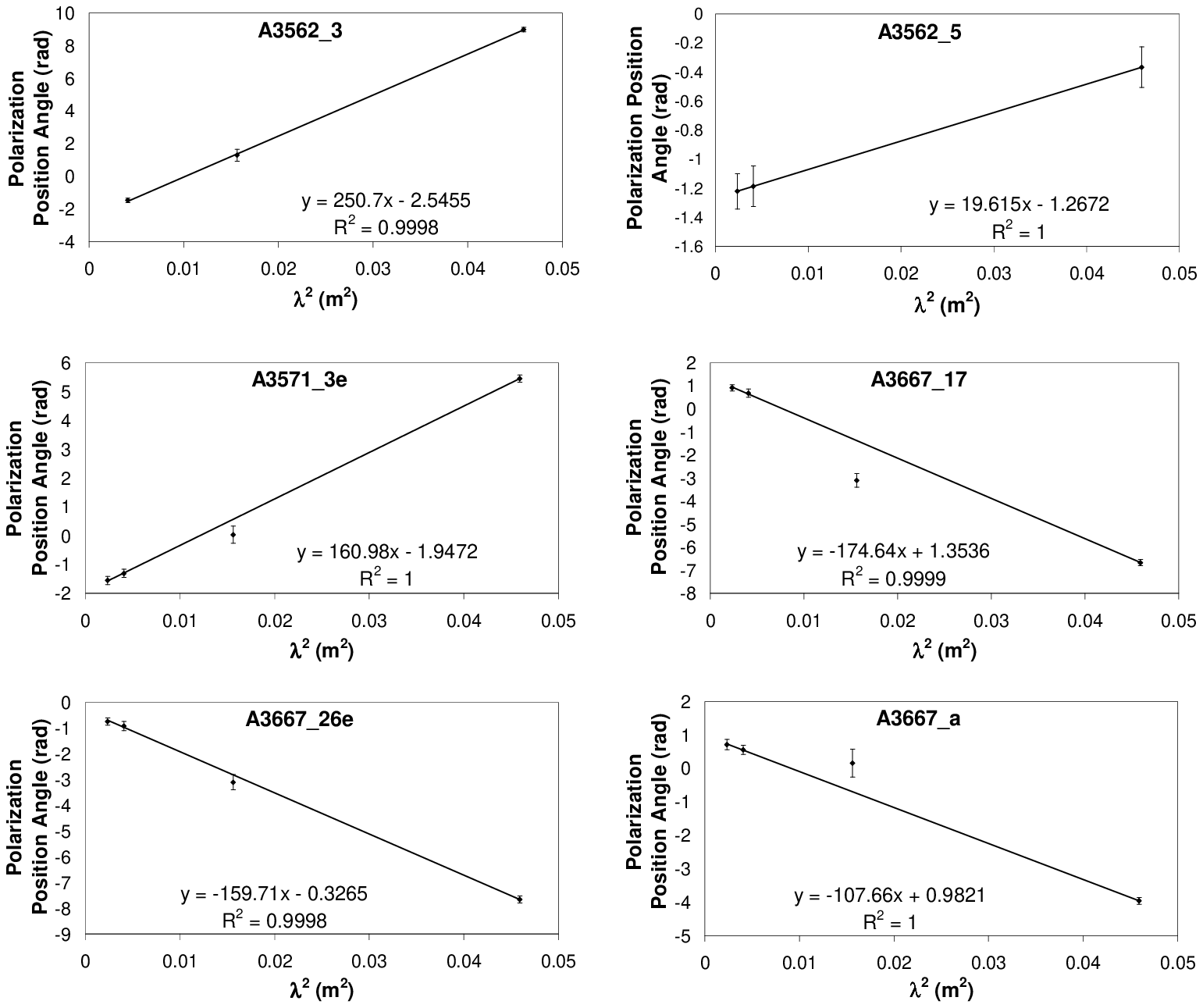}}
\resizebox{\hsize}{!}{\includegraphics{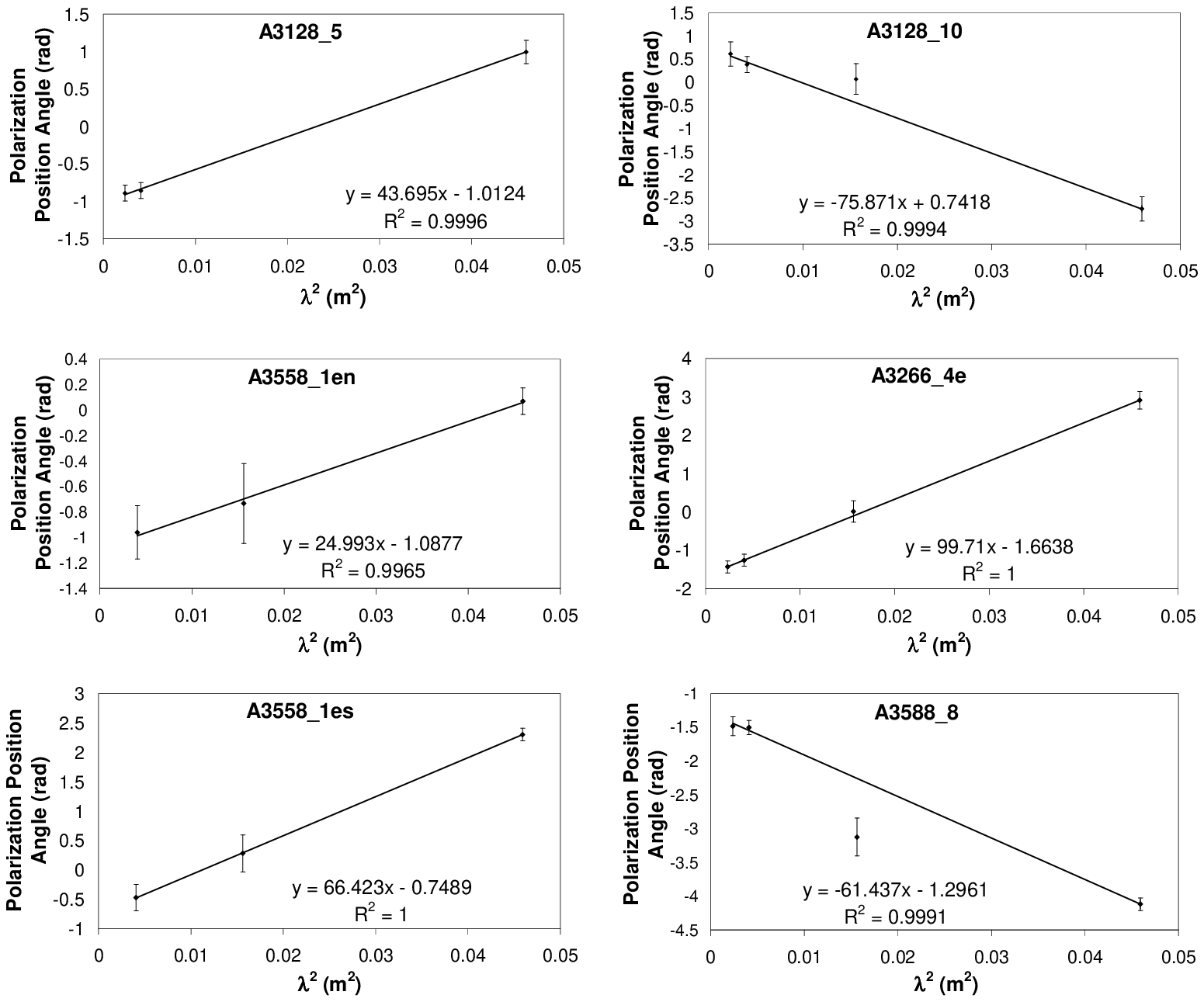}}
   \caption{RM fits for the cluster sample examined here. The graph shows the
equation of the line where the RM corresponds to the slope, the R$^{2}$ statistic  
which is a measure of the goodness of fit to a straight line is also 
given (an R$^{2}$ value of 1 corresponds to a perfect fit, the 13 cm data is given 
less weight in the fitting due to the poor ATCA off-axis response) .}
 \label{fig:rmfit}
\end{figure}

Plots of the position angle and frequency were examined. Due to the small
number of points to consider it was not necessary to write a complicated 
fitting routine to find the best fit. Data were adjusted by hand under 
the assumption that there would be no rotation between the two closely spaced points at 4.7 and
6.2 GHz and that a maximum of 360 degrees ambiguity was likely to 
have occurred in the 1.4 GHz value. Assuming no rotation between the closely 
space values at 4.7 and 6.2 GHz allows for $|$RM$|$ $\leq$ 1350 rad~m$^{-2}$ to 
be fitted unambiguously. The resultant points were then passed through a standard linear 
least-squares fitting routine. It was found in many cases that the 2.4 GHz points 
were difficult to reconcile with the other three measurements, giving slightly 
discrepant values. Despite efforts to minimise errors this is likely to be a 
result of the polarisation response of the 13cm feed on the ATCA. Though
sources were observed near the beam centre in order to have the best polarisation
characteristics at all frequencies it appears that some effect is still evident in the
13 cm data. Thus, the 13cm points were given less weighting
during the fitting procedure.
Figure \ref{fig:rmfit} shows the resultant plots, 
while the RMs obtained are listed in Table \ref{tab:theta}. 
Surprisingly the fitting worked extremely well with the 
worst case fit to the data still giving a 99 \% confidence to a straight line.

It should be noted that the use of observations from a period 
spanning 18 months gives some concern as the data are not all from the same
epoch. Nevertheless the quality of the RM fits is excellent and it is 
therefore assumed that there is no significant variation in the observed source RMs over 
this time scale.

\section{Results: Comparison to Other Data}
\label{results}

The RMs obtained from the fitting procedure were then corrected for
the contribution from the Galactic rotation measure, G$_{RM}$. This was done by
using an interpolated all-sky rotation measure map generated from published 
RM catalogues \citep{mjh04}. The Galactic contribution was subtracted 
from the measured RM to give a residual RM, (RRM) which represents a 
combination of the cluster RM and the intrinsic source RM. 
Previous studies have used the standard deviation of 
the distribution of extra-galactic RMs at high galactic latitudes,
beyond the influence of the Galaxy, to argue that the contribution from 
internal RMs is small. The standard deviation of around 400 extra-galactic RMs
at greater than 30 degrees from the Galactic plane was found to be 
10 rad~m$^{-2}$. This suggests that 
the intrinsic RM component should be small 
and that RRM should adequately represent the Cluster contribution to the 
measured RM. Previously it was assumed that the distribution was Gaussian and 
thus the likelihood of encountering a moderate to high intrinsic RM 
was very low. However, further analysis of the high Galactic 
latitude extra-galactic RM population has shown the distribution 
is exponential at above the 99.9\% confidence level. This means that it is
more likely  to observe a background source with a 
significant internal contribution to the measured RM than previously thought. 
This reinforces the requirement to examine cluster magnetic fields statistically.

\begin{figure}[htbp]
\centering
\resizebox{\hsize}{!}{\includegraphics{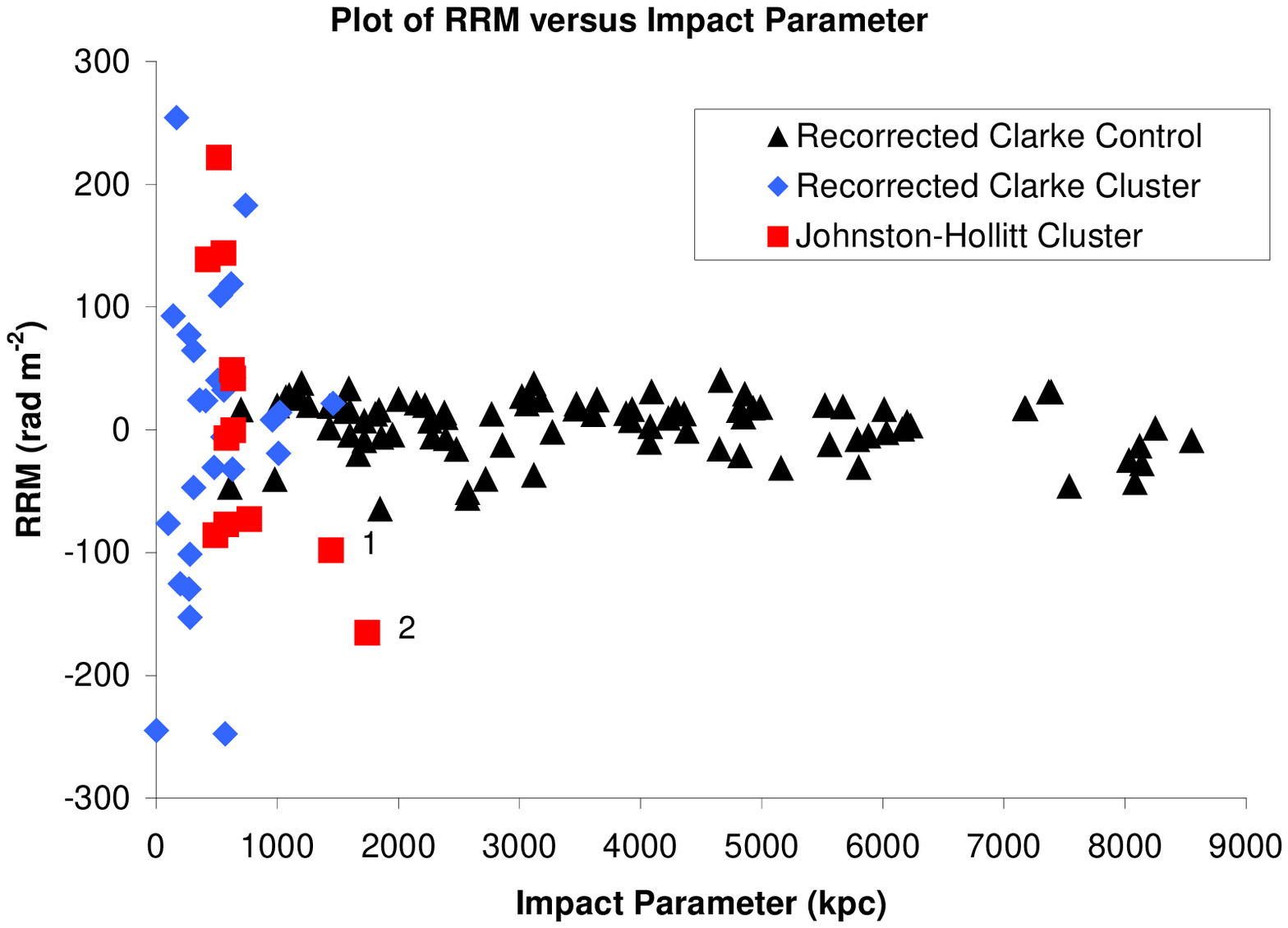}}
\resizebox{\hsize}{!}{\includegraphics{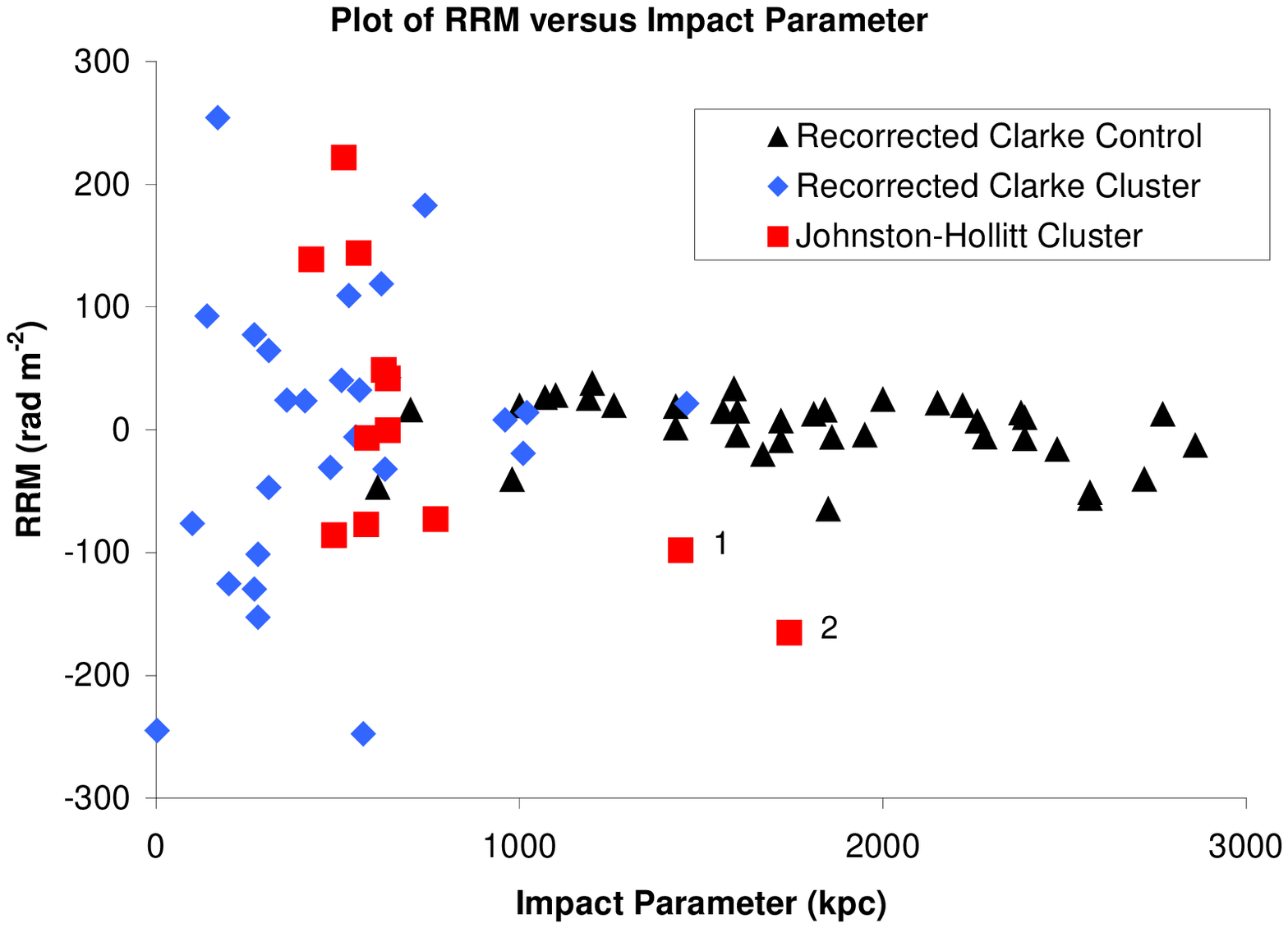}}
   \caption{Impact Parameter versus Residual Rotation Measure. The labelled points
correspond to sources seen in projection through a region of diffuse synchrotron
emission in A3667.}
 \label{fig:RMplot}
\end{figure}

\begin{figure}[htbp]
\centering
\resizebox{\hsize}{!}{\includegraphics{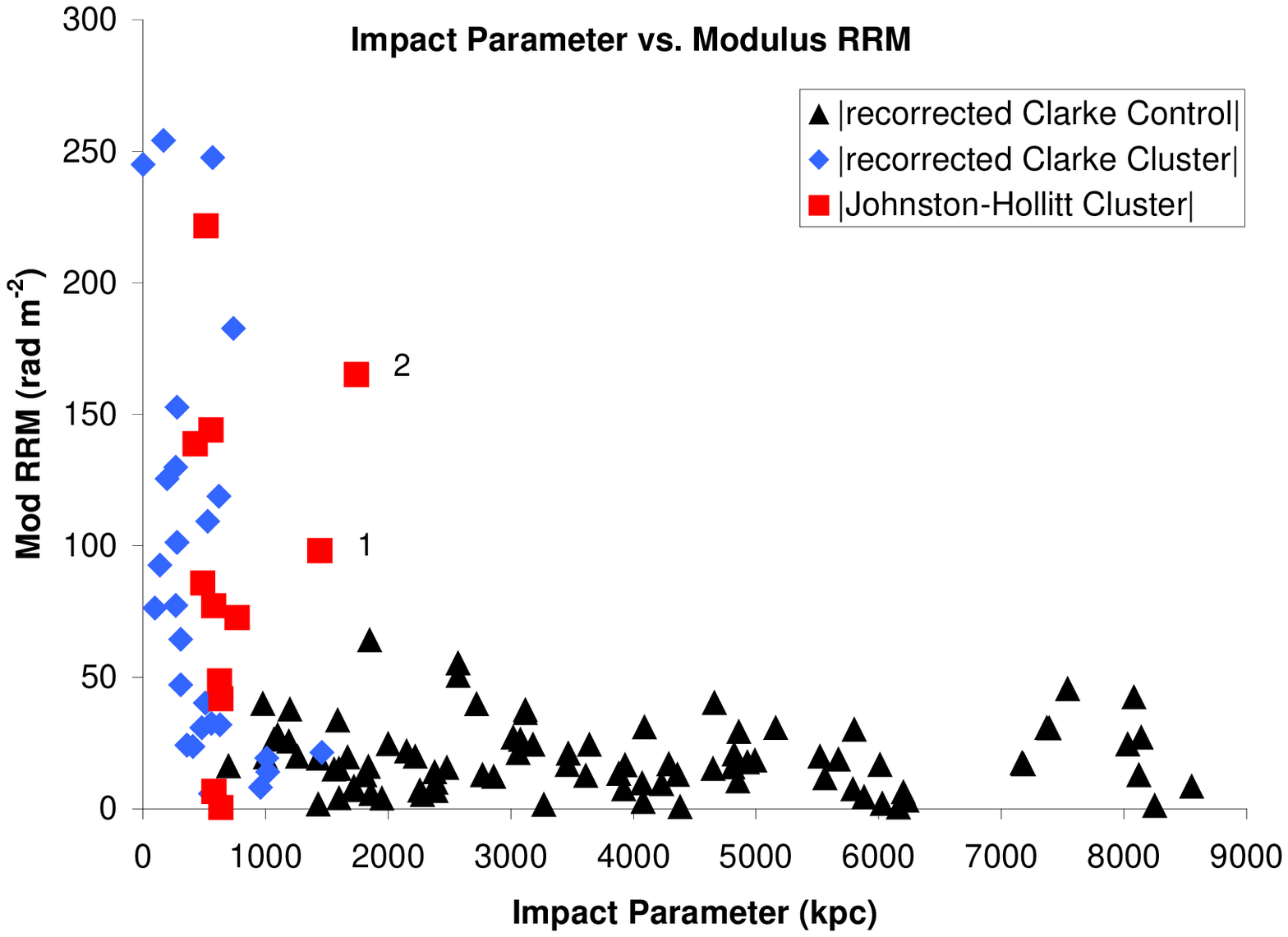}}
\resizebox{\hsize}{!}{\includegraphics{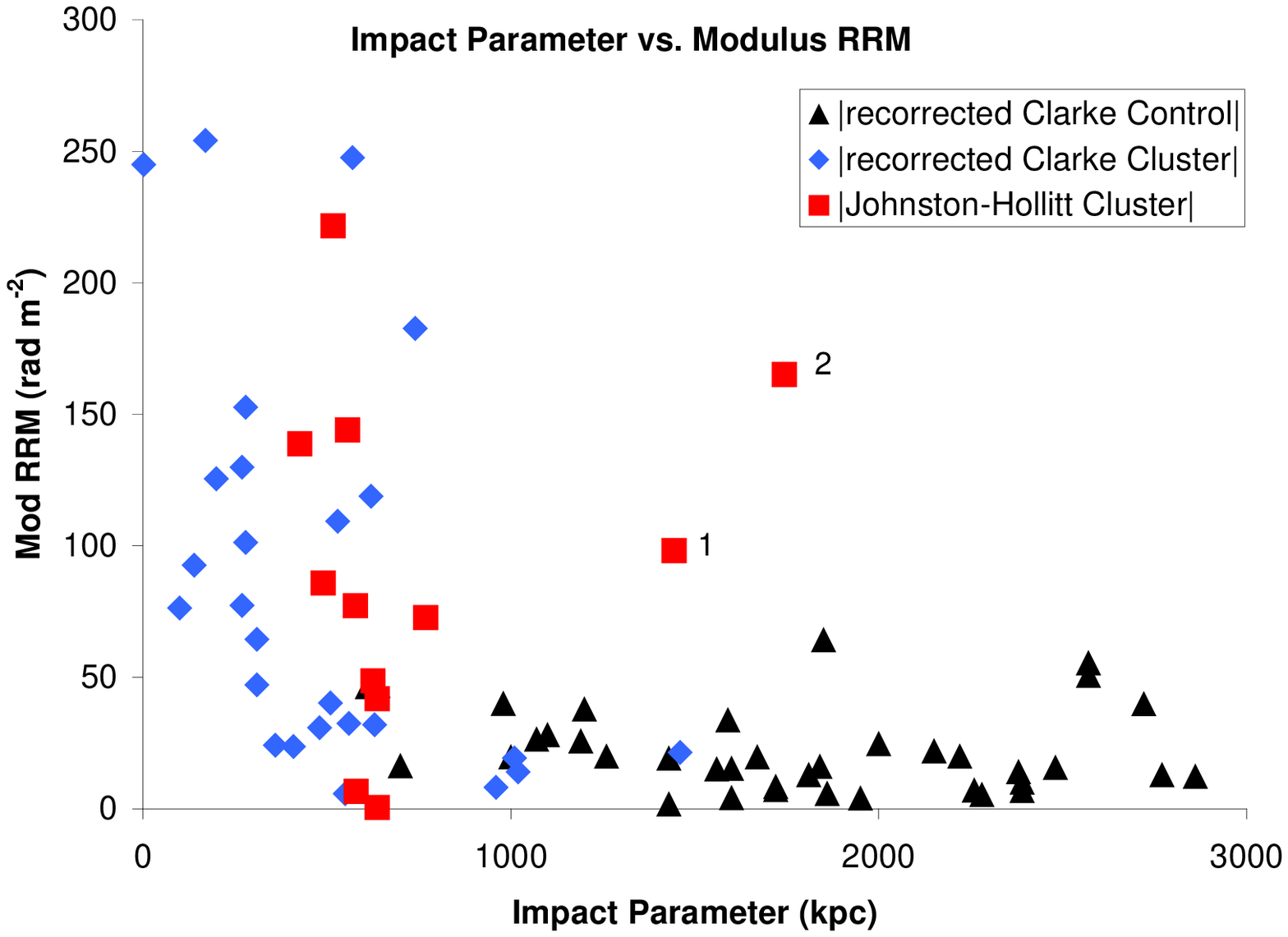}}
   \caption{Impact Parameter versus Residual Modulus Rotation Measure. The labelled points
correspond to sources seen in projection through a region of diffuse synchrotron 
emission in A3667.}
 \label{fig:RMplot2}
\end{figure}

\citet{Clarkephd} and \citet{Clarke01} also corrected for the Galactic 
contribution via examining published RMs in a 15 degree radius about each 
cluster. In order to directly compare the two datasets, the RRMs from the 
\citet{Clarkephd} sample were recalculated using the interpolated map. In most
cases this made a 5-10\% in the RRM values. The
two samples were then plotted together on two graphs showing distance from the
cluster centre (the so-called impact parameter) versus RRM and $|$RRM$|$. 
These plots are shown in Figures \ref{fig:RMplot} and \ref{fig:RMplot2} 
respectively.
Each plot shows the entire combined dataset out to an impact parameter of
9000 kpc in the top panel and restricted impact parameter range of 0 to 3000 
kpc in the lower panel. The RRM of the southern cluster 
sample presented here (see Table \ref{tab_RRM}) agrees well with the 
northern sample of Clarke which drops to a background level at around 
800 kpc from the cluster core.
The two labelled points are both from A3667 and are background sources to
the largest and brightest diffuse radio emission region yet discovered 
\citep{Rottgering97, mjh03}. These two points are significantly above the RM level 
suggested by the other data. They are further beyond the region of X-ray 
emission in A3667 and should fall at a background level, the fact that they do
not strongly suggests that these RMs are probing the magnetic field of the
diffuse radio emission.

The data for both lobes of the source A3558.1e show quite different RRMs 
(-0.5 and 41.9) despite being closely spaced. This may either be interpreted
as due to tangling of the cluster magnetic field on scales of order of 10 kpc,
or as a difference in the internal properties of the radio lobes. Observations
of radio jets in low power radio sources have shown that the magnetic 
field is aligned along the jet and becomes tangled in the resultant lobes due
to entrainment. Thus, it is likely here that we are seeing a combination internal
and environmental effects.

\begin{table}
\label{tab_RRM}

\caption{Residual RM corrections. Column 1 gives
the source name used in the ATCA observing program and archive; col 2 
is the J2000 galactic longitude; col 3 is the J2000
galactic latitude; col 4 is the Galactic contribution to the rotation measure
at these co-ordinates as calculated with an interpolated all-sky rotation
measure map \citep{mjh03b} and col 5 is the residual rotation measure for
each source once the Galactic contribution is subtracted.}

$$
\begin{array}{lllll}
\noalign{\smallskip}
\noalign{\hrule}
\noalign{\smallskip}
 {\rm Source}     &   {\rm l}   & {\rm b} & {\rm G_{RM}} & {\rm RRM}  \\
\noalign{\smallskip}
\noalign{\hrule}

  {\rm A}3266\_4{\rm e} & 272.29 & -40.23 & -44.3 & 144.0\\
  {\rm A}3128\_5  & 262.75 & -48.20 & -4.9 & 48.6\\
  {\rm A}3128\_10 & 264.82 & -50.91 & -3.2 & -72.7\\
  {\rm A}3558\_1{\rm en} & 312.15 &  30.88 &  24.5 & -0.5\\
  {\rm A}3558\_1{\rm es} & 312.15 &  30.88 &  24.5 & 41.9 \\
  {\rm A}3558\_8  & 312.11 &  30.62 &  24.5 &-85.9 \\
  {\rm A}3562\_3  & 313.37 &  30.50 &  29.1 & 221.6\\
  {\rm A}3562\_5  & 313.52 &  30.34 &  26.3 & -6.7\\
  {\rm A}3571\_3{\rm e} & 316.49 &  28.60 &  22.2 & 138.8\\
  {\rm A}3667\_{\rm A}  & 341.32 & -33.21 & -9.5 & -98.2\\
  {\rm A}3667\_26{\rm e}& 340.98 & -33.25 & -9.5 & -76.7\\
  {\rm A}3667\_17 & 341.19 & -32.97 & -9.5 & -165.1\\

\noalign{\hrule}
\end{array}
$$
\end{table}

A Kolmogorov-Smirnoff test to assess the likelihood that the combined cluster and the control 
samples are drawn from the same population was preformed. The test rejected the null hypothesis
at greater than 99 precent. 

\subsection{Embedded versus Background Sources}

Recently it has been claimed that the rotation measures of radio sources 
embedded in clusters are not useful as probes of the global cluster magnetic
fields but rather only probe the field local to the source \citep{larry}.
As a counter argument to this we have taken the subsample of those
galaxies presented here and in Clarke (2000) which are background to
the clusters and performed a Kolmogorov-Smirnoff (KS) test to assess if 
these data are drawn from the sample population as the control galaxies of Clarke (2000).
The KS test rejects the null hypothesis at greater than the 99\% confidence level
demonstrating that the two samples are not drawn from the same population. We also added the 
data from \citet{hoe} to the sample increasing both the cluster and control sample and again per
formed a
KS test to see if the two samples are drawn from the same population. Again the null hypothesis
 can be 
rejected at over 99\% confidence. In addition, we also
considered the hypothesis that the RMs derived from embedded and background sources might be
drawn from the same population. In this case the null hypothesis could not be rejected and it
seems, that statistically speaking at least, there is no difference in the value of RMs derived
 from
background or embedded cluster sources seen in project through galaxy clusters. Thus, 
despite concerns over the validity of the embedded 
galaxies as probes there is no statistical evidence this class of source gives rise to 
significantly different
RMs to background galaxies though both a combined embedded and cluster background sample gives 
rises to a significantly different RMs than the control sample. Further, even using only 
background sources there is a still statistically
significant excess RM detected along lines of sight through X-ray luminous clusters when 
compared with other lines of sight. With the exception 
of the points in A3667, our data agree well with the previous samples and support the 
finding by Clarke (2000) and show an excess RM toward the centre of galaxy clusters. We find 
the combined cluster RM sample of Clarke (2000), \citet{hoe} and these data has a standard deviation of 
$\sigma_{RM}$ = 125 rad~m$^{-2}$. In comparison, the standard deviation of a sample of 474 
extra-galactic sources at least 30 degrees from the galactic plane gave $\sigma$ =10 rad~m$^{-2}$ \citep{mjh03}.  

\section{Conclusion}
\label{conclusions} 
We have presented the results of a search to detected excess Faraday rotation toward the
cores of several southern, non-cooling flow clusters. We find that the population of RMs derived
from combined sample of data from this work and the literature through lines of sight through
galaxy clusters is statistically different to RMs from other lines of sight. Further, we find that
this holds even for a sample of only background galaxies seen in projection through clusters.
Additionally, a comparison of data from embedded and background sources can not reject the null
hypothesis that the values are drawn from the same population. In conclusion we argue that the 
results of this study agree well with those of Clarke (2000), supporting the notion that a 
statistically significant broadening of the RM distribution is measured out to
around 800 kpc for nearby galaxy clusters. This suggests cluster magnetic 
fields to be of the order of 1--2 $\mu$G assuming a tangled cell model.

\section*{Acknowledgements}
We thank Dr Tracy Clarke for providing updated information relating to her thesis work and her useful 
discussions on the topic. In addition, we thank Dr Larry Rudnick for interesting discussions in 
particular those on sources of error in statistical RM analyses. MJH extends her thanks to the 
staff of ATNF for their support during all stages of data collection and analysis. 
The Australia Telescope Compact Array telescope is part of the Australia Telescope 
which is funded by the Commonwealth of Australia for operation as a National Facility managed by CSIRO.
This research has made use of the NASA/IPAC Extragalactic Database (NED) which is 
operated by the Jet Propulsion Laboratory, California Institute of Technology, under 
contract with the National Aeronautics and Space Administration.

\bibliographystyle{aa}
\bibliography{paper}

\begin{thebibliography}{27}
\expandafter\ifx\csname natexlab\endcsname\relax\def\natexlab#1{#1}\fi

\bibitem[{{Broten} {et~al.}(1988){Broten}, {MacLeod}, \& {Vallee}}]{Broten}
{Broten}, N.~W., {MacLeod}, J.~M., \& {Vallee}, J.~P. 1988, \apss, 141, 303

\bibitem[{{Burbidge}(1958)}]{Burbidge}
{Burbidge}, G.~R. 1958, \apj, 128, 1

\bibitem[{{Burkard}(1961)}]{IonFRM}
{Burkard}, O. 1961, \jgr, 66, 3058

\bibitem[{{Clarke}(2000)}]{Clarkephd}
{Clarke}, T.~E. 2000, Ph.D.~Thesis

\bibitem[{{Clarke} {et~al.}(2001){Clarke}, {Kronberg}, \& {B{\"
  o}hringer}}]{Clarke01}
{Clarke}, T.~E., {Kronberg}, P.~P., \& {B{\" o}hringer}, H. 2001, \apjl, 547,
  L111

\bibitem[{{Cooper} \& {Price}(1962)}]{Cooper}
{Cooper}, B. F.~C. \& {Price}, R.~M. 1962, \nat, 195, 1084

\bibitem[{{Dennison}(1979)}]{d}
{Dennison}, B. 1979, \aj, 84, 725

\bibitem[{{Ebeling} {et~al.}(1996){Ebeling}, {Voges}, {Bohringer}, {Edge},
  {Huchra}, \& {Briel}}]{Ebeling96}
{Ebeling}, H., {Voges}, W., {Bohringer}, H., {et~al.} 1996, \mnras, 281, 799

\bibitem[{{Goldshmidt} \& {Rephaeli}(1993)}]{Goldshmidt93}
{Goldshmidt}, O. \& {Rephaeli}, Y. 1993, \apj, 411, 518

\bibitem[{{Hennessy} {et~al.}(1989){Hennessy}, {Owen}, \& {Eilek}}]{hoe}
{Hennessy}, G.~S., {Owen}, F.~N., \& {Eilek}, J.~A. 1989, \apj, 347, 144

\bibitem[{{Jaffe}(1980)}]{Jaffe}
{Jaffe}, W. 1980, \apj, 241, 925

\bibitem[{{Johnston-Hollitt}(2003)}]{mjh03}
{Johnston-Hollitt}, M. 2003, Ph.D.~Thesis

\bibitem[{{Johnston-Hollitt}(2004)}]{mjh03b}
{Johnston-Hollitt}, M. 2004, in The Riddle of Cooling Flows in Galaxies and
  Clusters of galaxies

\bibitem[{{Johnston-Hollitt} {et~al.}(2004){Johnston-Hollitt}, {Hollitt}, \&
  {Ekers}}]{mjh04}
{Johnston-Hollitt}, M., {Hollitt}, C.~P., \& {Ekers}, R.~D. 2004, in The
  Magnetized Interstellar Medium, 13--18

\bibitem[{{Kim}(1988)}]{Kimphd}
{Kim}, K. 1988, Ph.D.~Thesis

\bibitem[{{Kim} {et~al.}(1990){Kim}, {Kronberg}, {Dewdney}, \&
  {Landecker}}]{Kim90}
{Kim}, K.-T., {Kronberg}, P.~P., {Dewdney}, P.~E., \& {Landecker}, T.~L. 1990,
  \apj, 355, 29

\bibitem[{{Kim} {et~al.}(1991){Kim}, {Kronberg}, \& {Tribble}}]{ktk}
{Kim}, K.-T., {Kronberg}, P.~P., \& {Tribble}, P.~C. 1991, \apj, 379, 80

\bibitem[{{Kronberg} \& {Simard-Normandin}(1976)}]{Kronberg76}
{Kronberg}, P.~P. \& {Simard-Normandin}, M. 1976, \nat, 263, 653

\bibitem[{{Lawler} \& {Dennison}(1982)}]{ld}
{Lawler}, J.~M. \& {Dennison}, B. 1982, \apj, 252, 81

\bibitem[{{Rottgering} {et~al.}(1997){Rottgering}, {Wieringa}, {Hunstead}, \&
  {Ekers}}]{Rottgering97}
{Rottgering}, H.~J.~A., {Wieringa}, M.~H., {Hunstead}, R.~W., \& {Ekers}, R.~D.
  1997, \mnras, 290, 577

\bibitem[{{Rudnick} \& {Blundell}(2004)}]{larry}
{Rudnick}, L. \& {Blundell}, K.~M. 2004, in The Riddle of Cooling Flows in
  Galaxies and Clusters of galaxies

\bibitem[{{Sault} {et~al.}(1995){Sault}, {Teuben}, \& {Wright}}]{Sault95}
{Sault}, R.~J., {Teuben}, P.~J., \& {Wright}, M.~C.~H. 1995, in ASP Conf. Ser.
  77: Astronomical Data Analysis Software and Systems IV, 433--+

\bibitem[{{Simard-Normandin} {et~al.}(1981){Simard-Normandin}, {Kronberg}, \&
  {Button}}]{SN}
{Simard-Normandin}, M., {Kronberg}, P.~P., \& {Button}, S. 1981, \apjs, 45, 97

\bibitem[{{Taylor} {et~al.}(1994){Taylor}, {Barton}, \& {Ge}}]{Taylor}
{Taylor}, G.~B., {Barton}, E.~J., \& {Ge}, J. 1994, \aj, 107, 1942

\bibitem[{{Vallee}(1990)}]{Vallee90}
{Vallee}, J.~P. 1990, \apj, 360, 1

\bibitem[{{Vallee} {et~al.}(1986){Vallee}, {MacLeod}, \& {Broten}}]{Vallee86}
{Vallee}, J.~P., {MacLeod}, M.~J., \& {Broten}, N.~W. 1986, \aap, 156, 386

\bibitem[{{Welter} {et~al.}(1984){Welter}, {Perry}, \& {Kronberg}}]{Welter84}
{Welter}, G.~L., {Perry}, J.~J., \& {Kronberg}, P.~P. 1984, \apj, 279, 19

\end{thebibliography}

\end{document}